\documentclass
[aps,pre,showpacs,showkeys,superscriptaddress,twocolumn]{revtex4}%
\usepackage{amssymb}%
\usepackage{amsmath}%
\usepackage{amsfonts}%
\usepackage{graphicx}
\def\be{\begin{equation}}
\def\ee{\end{equation}}
\def\beq{\begin{eqnarray}}
\def\eeq{\end{eqnarray}}

\def\U{\bf U}
\def\W{\mathbf W}
\begin{document}
\title{ Preserving Quantum States Using Inverting Pulses  : A Super-Zeno 
Effect}
\author{Deepak Dhar}
\affiliation{Department of Theoretical Physics, Tata Institute of Fundamental Research,
Homi Bhabha Road, Mumbai 400005, India}
\author{L. K. Grover}
\affiliation{Bell Laboratories, Mountain Avenue, Murray Hill, NJ}
\author{S. M. Roy}
\affiliation{Department of Theoretical Physics, Tata Institute of Fundamental Research,
Homi Bhabha Road, Mumbai 400005, India}
\date{\today}

\begin{abstract}

We construct an algorithm for suppressing the transitions of a quantum mechanical
system, initially prepared in a subspace ${\mathcal{P}}$ of the full Hilbert space
of the system, to outside this subspace by subjecting it to a sequence of unequally
spaced short-duration pulses. Each pulse multiplies the amplitude of the vectors in
the subspace by $-1$. The number of pulses required by the algorithm to limit the
leakage probability to $\epsilon$ in time $T$  increases as  $T \exp[ \sqrt{
\log(T^2/\epsilon)}]$, compared to $T^2 \epsilon^{-1}$ in  the standard quantum 
Zeno effect.


\end{abstract}

\maketitle

Quantum computation is known to be more powerful than classical
computation. Two well-known examples where quantum computation gives a
clear advantage over the classical are searching and factorization. A key
difference is that, only in the quantum case, one can use the possibility
of destructive interference of quantum amplitudes to suitably design
algorithms where one can cancel out the amplitudes of undesired states.  
We want to apply this general strategy to the problem of state
preservation, i.e. how to limit the system's evolution to a desired
subspace \cite{ref1}.

Several different approaches to this question have been studied.
Error-avoiding codes \cite{ref2} depend on existence of subspaces free of
decoherence due to special symmetry properties, and error-correcting codes
\cite{ref3} depend on monitoring the system and conditional feedback to
achieve this. Another important strategy for state preservation is to use
the well known quantum Zeno effect, studied by von Neumann, and others
\cite{ref6}, viz. preserving a given quantum state by very frequent
measurements. In fact, the reduction of wavefuction is not essential for
the physics of the quantum Zeno effect, and one can achieve a similar
suppression of transitions by suitably coupling the system strongly to an
external system for short intervals in the bang-bang control \cite{ref4} 
and dynamical
decoupling strategies \cite{ref5}, where the time evolution is always
unitary, without any reduction of the wave-packet \cite{ref7}.

In this paper, we develop a bang-bang control type algorithm to preserve quantum
states that involves unitary kicks interspersed with evolution according to the
system-environment Hamiltonian for {\it unequal time-intervals} between between
kicks. The unitary kicks we employ are inverting pulses similar to those used in
Grover's quantum search algorithm \cite{grover}. The proposed algorithm does not
require any symmetry properties and is much more efficient than the quantum Zeno
effect.  We will show that in our algorithm, the number of pulses required to keep
the the quantum system in the same subspace up to time $T$ with probability greater
than $1-\epsilon$ increases only as $T \exp[ \sqrt{ \log( T^2/\epsilon)}]$, for
large $T$, and small $\epsilon$, whereas it varies as $T^2 \epsilon^{-1}$ and
$T^2 \epsilon^{-1/2}$ for the quantum zeno effect with and without measurement.
 We shall call the preservation of state using such
non-periodic pulse sequences the super-Zeno effect.

We consider a quantum mechanical system described by a finite dimensional
Hilbert space ${\mathcal{H}}$, which is a direct sum of two orthogonal
subspaces ${\mathcal{P}}$ and ${\mathcal{Q}}$. We assume that the system
Hamiltonian $\mathbf{H}$ is bounded. The unitary operator corresponding to
evolution for a time interval $t$ is given by $\mathbf{U}_{0} (t) =
\exp[-i\mathbf{H}t]$.  When the system is subjected to a very short
duration external fields pulse, we assume that the effect of the pulse can
be represented by a unitary operator ${\mathbf J}$. If the system is
initially in the state $|\psi\rangle$, after being subjected to the pulse,
its state is $\mathbf{J}|\psi\rangle$. 

A sequence of $N$ pulses is specified by $(N+1)$ real numbers $\{x_j\}, j
= 1 {\rm ~to~ } N+1$, where $x_j t$ is the time-interval between the $j$th
and $(j+1)$th pulse for $j= 1$ to $N-1$, and $x_{N+1} t$ is the interval 
between the last pulse, and the measurement of the  state of the system. 
Here
$t$ is time interval between the initial preparation, and final
measurement, and $ \sum_{j=1}^{N+1} x_j =1$.  In this case, the evolution
operator is

\beq
\W_N (t) = {\bf U}_0(x_{N+1}t) {\mathbf J}   \ldots  {\mathbf J}  {\bf 
U}_0(x_2 t) {\mathbf J}
{\bf U}_0(x_1 t).
\label{eqw}
\eeq

Our aim is to choose a sequence of the  time-intervals between
pulses given by $\{x_j\}$, such that if the system is initially in a state
$|\psi\rangle \in{\mathcal{P}}$, then the probability that
the system is found in a state in ${\mathcal{Q}}$ after the pulse sequence
is minimized.

In this paper, we shall discuss only the case where ${\mathbf J}$ is an
inverting pulse:  $\mathbf{J} = \mathbf{Q}-\mathbf{P}$, where
$\mathbf{P}$ and $\mathbf{Q}$ are the projection operators for the
subspaces ${\mathcal{P}}$ and ${\mathcal{Q}}$ respectively. Clearly,
${\mathbf J} |\psi\rangle = -|\psi \rangle$, if $|\psi \rangle \in
{\mathcal P}$, and $ {\mathbf J} |\psi\rangle = +|\psi \rangle$, if $|\psi
\rangle \in {\mathcal Q}$.  We first give an explicit construction of a recursively
defined sequence ${\mathbf U}_m(t)$ (see Eq. (4) below), such that the transition amplitude
$\langle q| {\mathbf U}_m(t)|p\rangle$ is ${\mathcal O}(t^{m+1})$. We then 
obtain quantitative bounds
on the leakage probability, and compare the performance of our algorithm with the standard quantum Zeno 
effect.

The basic idea of using inverting pulses to produce destructive
interference of quantum mechanical amplitudes and reduce the transition
rate from a particular subspace to others is quite straight-forward.  Let $|p\rangle$ be a general 
state in ${\mathcal{P}}$, and
$|q\rangle$ a general state in ${\mathcal{Q}}$. Let ${\mathbf U}(t) $ be a
unitary operator satisfying $\mathbf{U}(t) = \mathbf{I} + O(t)$ for $t
\rightarrow 0$, where \textbf{I} is the identity operator and the matrix
elements $U_{qp}$, and $U_{pq}$ are of $O(t^{r})$, with $r$ is a positive
integer. Then in the transition amplitude $\langle
q|\mathbf{U}(t)\mathbf{J} \mathbf{U}(t)|p\rangle$ a  precise
cancellation of the amplitude of ${\mathcal O}(t^r)$ term occurs, and one
gets \begin{equation} \langle q|\mathbf{U}(t)\mathbf{J}
\mathbf{U}(t)|p\rangle= O(t^{r+1}). \label{eq3} \end{equation} Similarly,
it is easy to see that if ${\mathbf V}(t)$ is an operator with 
$\mathbf{V}(t) = \mathbf{J} + O(t)$, then again a
destructive interference of amplitudes to leading order occurs in $\langle
q |\mathbf{V}(t)^{2}| p\rangle$, and if $V(t)_{qp} = O(t^{r})$, then,
\begin{equation} \langle q |\mathbf{V}(t)^{2}| p\rangle= O(t^{r+1}).
\label{eq4} \end{equation}

 Using this result, it is straight forward to construct a pulse sequence
where the transition amplitude is ${\mathcal O}(t^r)$ for any  positive 
integer $r$ by
recursion. We define operators $\mathbf{U}_{m}$ by the recursion relations
\begin{eqnarray}
{\bf U}_{m+1}(t) &=& {\bf U}_{m}(t/2) ~{\mathbf J} ~{\bf U}_{m}(t/2), {\rm 
~for~} 
m 
{\rm~ even}, 
\nonumber \\
& =& {\bf U}_{m}(t/2) ~{\bf U}_{m}(t/2),{\rm ~for} ~m  {\rm ~odd.}
\end{eqnarray}
with ${\bf U}_{0} = {\bf U}_{0}(t)$. Then, by induction, it follows that  the transition amplitude 
$\langle q|{\bf U}_m|p\rangle$ is of order ${\mathcal O}(t^{m+1})$.

We can write down $\mathbf{U}_{m}(t)$ explicitly as a product of 
$\mathbf{U}_{0}(t/2^{m})%
$'s and $\mathbf{J}$'s. For example
\begin{eqnarray}
\mathbf{U}_{1}(t)   = \mathbf{U}_{0}(t/2) {\mathbf J}
\mathbf{U}_{0}(t/2), \nonumber\\
\mathbf{U}_{2}(t)   = [\mathbf{U}_{0}(t/4) {\mathbf J}
\mathbf{U}_{0}(t/4)]^{2},\nonumber \\
\mathbf{U}_{3}(t)   = 
[\mathbf{U}_{0}(t/8) {\mathbf J}
\mathbf{U}
_{0}(t/8)]^{2} {\mathbf J} [\mathbf{U}_{0}(t/8) {\mathbf J} 
\mathbf{U}_{0}(t/8)]^{2}. 
\end{eqnarray}

Let $N_m$ be the  number of pulses 
used  the sequence ${\mathbf U}_m$, then it is easily verified  that
\begin{eqnarray}
N_{m}=(2^{m+1}-2)/3, {\rm ~for }~m  {\rm ~even},\nonumber\\
     =(2^{m+1}-1)/3, {\rm ~for}~ m {\rm ~odd} .\label{twentynine}%
\end{eqnarray}

We can derive quantitative bounds to estimate the probability of leakage 
out
of the subspace ${\mathcal{P}}$ for this sequence $\mathbf{U}_{m}$. We 
will show that the leakage probability $L_{m}$ from subspace $%
\mathcal{P}$ to subspace $\mathcal{Q}$ for operator $\mathbf{U}_{m}$ 
applied for a total  time $T$ is
\begin{equation}
L_{m} = \sum_{q \in Q} |\langle q|\mathbf{U}_{m}|p\rangle|^{2} \leq
2^{-m(m+1)} [ET]^{2m+2}.  \label{eq10}
\end{equation}
Here $E = |\mathbf{H}|$, and we have used the definition of
the norm of an operator 
$\mathbf{A}$, given by
\begin{equation}
|\mathbf{A}| = {%
%TCIMACRO{\QATOP{_{_{}\mathrm{Sup}} }{\psi}}%
%BeginExpansion
\genfrac{}{}{0pt}{}{_{}\mathrm{ Sup}} {\psi}%
%EndExpansion
} \ ||\mathbf{A}|\psi\rangle||~/~||~|\psi\rangle||%
\end{equation}
By making $m$ large, $L_m$ can be made as small as we 
please. 
This becomes a state preservation algorithm when the subspace 
${\mathcal{P}}$
consists of just the initial state. Note that  the minimum time interval between pulses
decreases as $2^{-m} T$, and this may limit the maximum value of  
$m$ that can be realized in an experimental setup.

For the $m$-th sequence, , the leakage probability varies as $2^{-m^2}$,
while the number of pulses increases as $2^m$. Thus, the leakage
probability varies as $\exp[ -(\log N)^2]$.  In Fig. 1, we have plotted
the upper bound on  leakage probability $L_m$  as a function of the number 
of pulses $N_m$. For comparison, we have also shown the corresponding 
bound on the  leakage probability for same number of pulses
for the quantum Zeno effect, both with repeated measurements, and with
unitary evolution with periodic inversion pulses. For the last two, these 
bounds are  $(ET)^2/N$ and $(ET)^4/N^2$ rspectively ( see
discussion below). The improvement in the super-Zeno over the others is
quite large, even for fairly small values of $m$; e.g. for $E T =1$ and
$m=4$, which corresponds to only $10$ pulses, the leakage probability is
already of order $10^{-4}$.

\begin{figure}
\begin{center}
\includegraphics[width=5.0cm,height= 8cm,angle=-90]{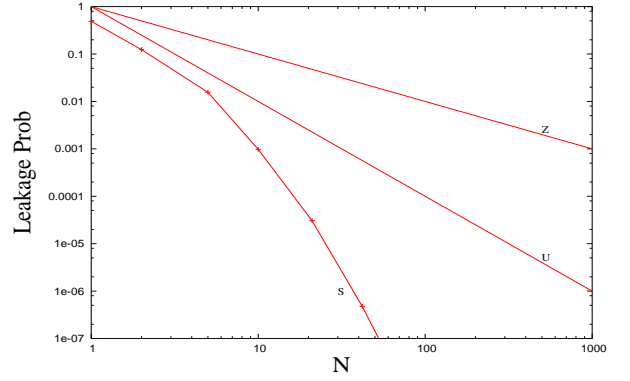}
\caption{The plot of the upper bounds to the leakage probability versus 
the number of pulses. The 
curves are marked Z   for the Zeno with periodic 
measurement, U for unitary evolution with periodic pulses, and S for 
the Super-Zeno cases, for $ E T = 1$. Here $T$ is the total time for evolution, and 
 $E$ is the norm of the hamiltonian.  }
\end{center}
\end{figure}

To prove the inequality (\ref{eq10}),  define
\begin{equation}
B_{m}(t) =\genfrac{}{}{0pt}{}{_{}{Max} }{|p\rangle |q\rangle}~~ |\langle 
q|\mathbf{U}_m(t)|p\rangle| = |\mathbf{Q} \mathbf{U}_m
\mathbf{P}|
\end{equation}

We have already seen that $B_m$ is ${\mathcal O}(t^{m+1})$. Since  the 
only available timescale is  $1/|{\mathbf H}|$, and $B_m$ 
is bounded by $1$ for large $t$, we can choose constant $K_m$ such that 
\beq 
B_m(t) \leq K_m (Et)^{m+1},{\rm ~for~~all }~t.
\label{eq10a}
\eeq
The proof of Eq.(\ref{eq10}) is complete if we can show that
\begin{equation}
K_m \leq 2^{-m(m+1)/2}.
\label{xx}
\end{equation}
This can be proved using the following 
simple mathematical inequalities  valid for all mutually orthogonal
projection operators $\mathbf{P}$ and
$\mathbf{Q}$,  and all unitary operators $\mathbf{U}$ and
$\mathbf{V}$, with ${\mathbf J} = {\bf Q} - {\bf P}$  (proof  omitted):
\begin{equation}
|\mathbf{Q} \mathbf{U} \mathbf{J} \mathbf{U}
\mathbf{P}| \leq~2~ |\mathbf{Q} \mathbf{U}
\mathbf{P}|~ \mathrm{Min}[1,~|\mathbf{U -I}|~ 
],\label{eq11}%
\end{equation}
and
\begin{equation}
|\mathbf{Q} \mathbf{V V} \mathbf{P}| \leq~2
~|\mathbf{Q} \mathbf{V} \mathbf{P}|~
\mathrm{Min}[1,~|\mathbf{V} -\mathbf{J}|~].\label{eq12}%
\end{equation}

Putting  $\mathbf{U  = U}_{r-1}$, with $r$ odd, and ${\mathbf V} ={\mathbf 
U}_{r-1}$ 
with $r$ even, then using $|{\mathbf U} -{\mathbf I}| 
\leq Et$ and $|{\mathbf V} -{\mathbf  J}| \leq Et$   in the inequalities 
above, we get
\beq
K_{r} \leq 2^{-r} K_{r-1}.
\eeq 
This, together with $K_0 =1$, implies Eq.(\ref{xx}), and completes the proof of Eq.(\ref{eq10}).

We now compare the performance of our algorithm with that of state
preservation using the standard Zeno effect. Let $C(T,\epsilon)$ be the
minimum number of pulses required to keep a quatum system in a precribed
state for time $T$, with the leakage probability less than $\epsilon$. It
seems reasonable to take this as the measure of the cost of algorithm.

Recall that in the standard quantum Zeno effect, starting with an initial
state $|p\rangle$, and measuring the projector $|p\rangle\langle p|$
repeatedly, at times $T/N$, $2T/N$, $\cdots$, $T$, the probability of
finding the system in the state $|p\rangle$ in each of these measurements
is $S(T)= |\langle p|e^{-i\mathbf{H}T/N}|p\rangle|^{2N}$. For small $T/N$,
we can approximate this as $ [ 1 - A (T/N)^2]^N \leq (1 - A T^2/N)$, where
$A= \langle p| {\bf H}^2 | p\rangle -\langle p| {\bf H}|p\rangle^2 \leq
E^2$. Thus the leakage probability is has an upper bound $(ET)^2/N$, and
$C(T,\epsilon)$ varies as $ T^{2}/\epsilon$.

If we use $N$ equispaced unitary kicks ${\mathbf J}$ in time $T$, the cost
$C(T,\epsilon)$ varies as $T^2/\sqrt{\epsilon}$. This follows from the
fact that the magnitude of the transition amplitude from subspace
${\mathcal P}$ to ${\mathcal Q}$ after two pulses is $ \leq (ET)^2/N^2)$,
and , the magnitude of the transition amplitude after $N$ pulses, and not
the transition probability, has an upper bound $(ET)^2/N$.

In our case, the number of pulses $N_m$ varies as $2^m$, and the leakage
probability varies as $\exp(- m^2)$, for large $m$.  Hence, for fixed $T$
and large $m$, the leakage probability $L_{m}$ decreases as $2^{ - (\log_2
N)^2}$. Equivalently, for fixed $T$,as $\epsilon$ is decreased,
$C(T,\epsilon)$ increases as $\exp( K \sqrt{ \log( 1/{\epsilon})})$, where
$K$ is a constant. This is slower than any power law in $\epsilon$.

Now we discuss the behavior of $C(T,\epsilon)$ for large $T$. In this 
case,  Eq.(\ref{eq10}) implies that $C(T,\epsilon)$ increases at most as 
$T^2$ for 
large $T$. However, for large $T$,  the bound  Eq.(\ref{eq10}) is too 
weak, and we can prove the  stronger result 
\beq
C(T,\epsilon) \leq K ET~ 2^{\sqrt{ \log_2 (E^2 T^2/\epsilon)}},
\label{xy}
\eeq
where $K$ is some constant.

    We indicate the proof here. Consider $2^{r+1} > ET >  2^{r}$, where 
$r$
is a non-negative integer. Assume $m \geq r$.  From the 
first halves of 
inequalities (\ref{eq11}) and (\ref{eq12}),   $B_m(T) \leq 
2^r B_{m-r}(T/2^r)$. Then using
the bound (\ref{eq10}) for $B_{m-r}$, we get
\begin{equation}
B_m(T)  \leq 2^r  (ET 2^{-m})^{m-r+1} 2^{\frac{(m-r)(m-r+1)}{2}}.
\end{equation}
On taking logs, the inequality  $ B_m \leq  \sqrt{\epsilon}$, becomes a 
quadratic inequality for $m$, which is easily solved to give 
\be
m \geq \log_2 (ET) + \sqrt{ \log_2( E^2 T^2/ \epsilon)} +{\mathcal O}(1),
\ee
for $ET \gg 1$. Since the number of pulses increases as $2^m$, we get 
Eq.(\ref{xy}).

Our  result is similar to that found by Khodjasteh and Lidar
\cite{khodjasteh}, who also used a recursive construction of pulse
sequences, and found it to be better than the periodic pulse sequence.  
They considered pulse sequences for preservation of quantum states which
work for all initial states, and the pulse sequence is independent of the
state to be preserved.  However, they require more than one type of pulses
to achieve this.  In contrast, we allow only one type of pulses, and can
only minimize the leakage probability out of a subspace, and the pulse
depends on the subspace to be preserved.

We note that in our construction, the intervals between pulses take only
two values, one of which is twice the other. This may be convenient in
actual implementation, as one can start with a source that generates
periodic field pulses of the desired type,  and then selectively block
some of the pulses. There is no problem in relaxing the assumption about
pulse-width being infinitesimal. In fact, if the inverting pulse is
error-free, then one can just use a periodic repetition of inverting
pulses to get a perfect preservation of the state. Effect of noise in the
inverting pulse would strongly affect the performance of the algorithm,
and remains to be studied.

It is possible to decrease the required number of pulses significantly if
we allow the intervals between pulses to be varied continuously
independent of each other.  In general,  one can try
to find non-negative $(N+1)$ real numbers $\{x_j\}$, such that for the 
matrix ${\mathbf W}_N(t)$ defined  in Eq.(1),
the transition amplitude $\langle q| \W_N(t) | p\rangle$ is of a
specified order $\mathcal{O} (t^{m+1})$, as $t \rightarrow 0$.  This is
certainly possible, by our explicit construction, for $N= N_m$.  Also, if
this is possible to do for one value of $N= r$, it is also possible for
$N=r+1$, as one can always add the $(r+1)$-th pulse at the end of time
$T$, with $x_{r+2}=0$. Hence we have proved that for any $m \geq 0$, there
is an integer $N_{min}(m)$, such that for all $N \geq N_{min}(m)$, one can
find an $N$- pulse sequence $\W_N(t)$ such that the transition
amplitude $\langle q| \W_N(t) | p\rangle$ is $\mathcal{O}
(t^{m+1})$, for small $t$.

Clearly, $N_{min}(m) \leq N_m$. We can improve this bound  using 
a
variation of the recursive construction of reflection-symmetric sequences
( i.e. those for which $x_{j}=x_{N+2-j}$, for all $j$) due to
Yoshida\cite{yoshida}. For such sequences, we have $\W_N(t) \W_N(-t)  =
{\mathbf I}$ for all $t$.  This implies that if $\langle
q|\W_N(t)|p\rangle = {\mathcal O}(t^{r})$, then $r$ must be odd.  Then
construct \beq \W'(t) = \W_N (\alpha t) {\mathbf J} \W_N ( \beta t)
{\mathbf J} \W_N (\alpha t), \eeq with $\alpha = 1/( 2 + 2^{\frac{1}{r}}),
\beta = 1 - 2 \alpha$. Then it is easily checked that the ${\mathcal
O}(t^{r})$ term in $\langle q | \W' |p \rangle$ vanishes by construction.
Hence the leading term in $\langle q | \W' |p \rangle$ must be the next
odd term, i.e. ${\mathcal O}(t^{r+2})$.  Starting with number of pulses
$N_r = 0$ for $r =1$, we can recursively construct a sequence where the
transition amplitude is ${\mathcal O}(t^{2 m +1})$ uses only $3^m - 1$
pulses. Thus, we get $N_{min}(2 m)  \leq 3^m -1$, which is a significant 
improvement over $N_{min}(m) \leq N_m$.

Further improvements are posible.
For a given $N$, one can express $\langle q|{\bf 
W}_N(t)|p \rangle$ as Taylor series
in powers of $t$. The coefficients of different powers of $t$ are sums of
matrix-elements of the type $\langle q| {\bf H}^{n_1} {\mathbf J} {\bf
H}^{n_2}{\mathbf J} ..|p \rangle$, with coefficients that are polynomials
of $\{x_j\}$. We try to set all such coefficients for powers of $t$ up to
$m$ equal to zero, and solve the resulting polynomial equations for
$\{x_j\}$. This calculation is straight forward for small $m$. For $m =0,
1$ and $2$, the sequences we get are $ \U_0, \U_1$ and $\U_2$. For larger
$m$, $\U_m$ or the ${\bf W}'$ are not optimal.

The number of polynomial equation for equating to zero all terms up 
to order $m$ are $(2^m -1)$.  
Interestingly, one can find a much small number of variables $\{x_j\} $
that do satisfy all of these. In particular, for $m=5$, we can find six
positive real numbers $x_1, \ldots x_6$ that satisfy all the $31$
equations generated. For still higher $m$, the equations become rather
messy to analyse, even with Mathematica.  However, in the cases that we
could solve, the final solution is surprizing simple.  We omit the
details, only mention here that for $m = 4$, the optimizing pulse sequence
is $\{x_j\} = \{ \beta, 1/4, 1/2 - 2 \beta, 1/4, \beta\}$ with $ \beta =
(3 - \sqrt{5})/8$. For $m=5$, it is $ \{\gamma, 1/4 -\gamma, 1/4, 1/4, 1/4
-\gamma, \gamma\}$, with $ \gamma = (\sqrt{3} -1)/4$. We find that while
$N_{min}(m)=m $ for $0\leq m\leq 5$, but $N_{min}(6) > 6$.

We note that the optimal sequences $\{x_j\}$ have been defined in terms of
${\bf H}$ and ${\mathbf J}$, but the polynomial equations, and therefore
the roots, actually do not depend on them at all! A more direct
characterization of these has not been possible so far.  Similar sequences 
have been studied earlier in the context of finding efficient numerical 
integration techniques in classical mechanics using symplectic integration
\cite{donnelly}. In quantum Monte Carlo studies,  the hamiltonian 
$\mathbf{H}$ of the system
can often be expressed as a sum of two mutually non-commuting terms :
$\mathbf{H} = \mathbf{A +B}$,  such that there are efficient procedures to 
apply the operators
$e^{t{\bf A}}$ and $e^{t{\bf B}}$ to a given state. Then using a
generalization of the well-known Trotter formula, the propagator of the
full hamiltonian is approximated as 
\beq 
e^{ t \mathbf{A} + t {\mathbf B}} = e^{x_1 t A} e^{x_2 t B} e^{x_3 t A} 
\ldots e^{x_N t B} + \mathcal{O}(t^m). 
\label{eq20} 
\eeq 

To make the discretization error small, one wants to choose the parameters
$x_1,x_2\ldots$ to maximize $m$ \cite{suzuki}. If we write $-i \mathbf{ H
= A}$ and $-i \mathbf{ J H J = B}$, then ${\mathbf J} e^{-i t x_j {\mathbf
H}} {\mathbf J} = e^{t x_j {\mathbf B}}$, and ${\mathbf W}_N(t)$ is of the
form Eq(\ref{eq20}). In the quantum Monte Carlo case, it has been proved
that for $m \geq 4$, one cannot find a solution with all $\{x_j\}$
non-negative \cite{suzuki2}.  In our problem, the inversion pulse provides
some crucial changes of sign in the algebraic equations, making the
construction of pulse sequences using non-negative $\{x_j\}$'s possible
for arbitrary $m$.

To summarize, in this paper, we have presented an algorithm for preserving
quantum states using inverting pulses, with unequal time interval betwen
pulses, which does not assume any symmetry properties of the Hamiltonian.  
The significant improvement over the quantum Zeno effect leakage
probability seems promising for applications to efficient preservation of
quantum states against decoherence, in quantum computations and
communications. On the theoretical side, the asymptotic behavior of
$N_{min}(m)$ for large $m$ is an interesting problem for further study.

DD would like to thank Prof. V. Singh and Prof. D.-N. Verma for very
enlightening discussions, and Sumedha for help with Mathematica. We thank
the referees, and K. Damle,  R. Godbole, T.R. Ramadas and M. Randeria for
their suggestions for improvement of the presentation of the paper.

\end{document}